\pdfoutput=1
\documentclass[sigconf]{acmart}
\usepackage{microtype}
\usepackage{mathtools}
\usepackage{tikz}
\usetikzlibrary{positioning}
\DeclarePairedDelimiter{\abs}{\lvert}{\rvert}
\DeclareMathOperator{\PSD}{PSD}
\begin{document}

\title[SAT+CAS: A Powerful Combination for Mathematics]{SAT Solvers and Computer Algebra Systems: \mbox{A Powerful Combination} for Mathematics}

\author{Curtis Bright}
\orcid{0000-0002-0462-625X}
\affiliation{%
  \institution{University of Waterloo}
  \streetaddress{200 University Ave W}
  \city{Waterloo}
  \state{Ontario}
  \country{Canada}
}
\email{cbright@uwaterloo.ca}

\author{Ilias Kotsireas}
\orcid{0000-0003-2126-8383}
\affiliation{%
  \institution{Wilfrid Laurier University}
  \streetaddress{75 University Ave W}
  \city{Waterloo}
  \state{Ontario}
  \country{Canada}
}
\email{ikotsire@uwaterloo.ca}

\author{Vijay Ganesh}
\orcid{0000-0002-8709-9351}
\affiliation{%
  \institution{University of Waterloo}
  \streetaddress{200 University Ave W}
  \city{Waterloo}
  \state{Ontario}
  \country{Canada}
}
\email{vganesh@uwaterloo.ca}

\let\para\paragraph
\renewcommand{\paragraph}[1]{\para{\bf#1}}

\acmConference{CASCON'19}{November 4--6, 2019}{Toronto, Ontario, Canada}
\acmYear{2019}

\begin{abstract}
Over the last few decades, many distinct lines of research aimed at automating
mathematics have been developed, including computer algebra systems (CASs) for
mathematical modelling, automated theorem provers for first-order logic,
SAT/SMT solvers aimed at program verification, and higher-order proof
assistants for checking mathematical proofs. More recently, some of these lines
of research have started to converge in complementary ways. One success story
is the combination of SAT solvers and CASs (SAT+CAS) aimed at resolving
mathematical conjectures.

Many conjectures in pure and applied mathematics are not amenable to
traditional proof methods. Instead, they are best addressed via computational
methods that involve very large combinatorial search spaces. SAT solvers are
powerful methods to search through such large combinatorial spaces---%
consequently, many problems from a variety of mathematical domains have been
reduced to SAT in an attempt to resolve them. However, solvers traditionally
lack deep repositories of mathematical domain knowledge that can be crucial to
pruning such large search spaces. By contrast, CASs are deep repositories of
mathematical knowledge but lack efficient general search capabilities. By
combining the search power of SAT with the deep mathematical knowledge in CASs
we can solve many problems in mathematics that no other known methods seem
capable of solving.

We demonstrate the success of the SAT+CAS paradigm by highlighting many
conjectures that have been disproven, verified, or partially verified
using our tool MathCheck.
These successes indicate that the paradigm is positioned to become a standard
method for solving problems requiring both a significant amount of search and
deep mathematical reasoning.  For example, the SAT+CAS paradigm has recently
been used by Heule, Kauers, and Seidl to find many new algorithms for
$3\times3$ matrix multiplication.
\end{abstract}

\maketitle

\section{Introduction}

The development of computer science 
has transformed the practice of mathematics.
The practical algorithms designed by computer scientists have
profoundly changed how many mathematical conjectures are proposed,
studied, and resolved.
For example, the fields of
\emph{satisfiability checking} and \emph{symbolic computation}
have each been paradigm-shifting in this way.  They have allowed
mathematicians the ability to solve problems much larger than
ever dreamt of in the past, the ability to pose and solve
entirely new kinds of mathematical conjectures, and the ability to
verify their solutions to unprecedented levels.

Despite a common background and over a hundred years of combined successful progress,
these two fields have developed mostly
independently of each other and have little common overlap~\cite{abraham2015building}.
It is in the interest of the working mathematician or computer scientist
to have familiarity with the techniques of these fields, as they
have broad (and often surprising) applicability.
This article provides an overview of these fields with an emphasis on how the
techniques of each field have been applied to resolve
mathematical conjectures---and how combining the techniques of each field
has resolved conjectures and solved problems that were out of reach of both fields.

\paragraph{Satisfiability checking}
The Boolean satisfiability (SAT) problem asks if it is possible
assign the variables in a Boolean logic expression in such a way
that the expression becomes true.  In the 1970s, the Cook--Levin
theorem demonstrated that the SAT problem is NP-complete resulting
in a pessimism that SAT problems are infeasible to solve in practice.
Despite this, research in the engineering of SAT solvers has
discovered algorithms and heuristics capable of solving enormous SAT instances
that cannot currently be solved by any other method.  This
``SAT revolution'' has had dramatic consequences for hardware and
software designers who now use SAT solvers on a daily basis~\cite{vardi2014boolean}.

In fact, SAT solvers have become so successful
that Heule, Kullmann, and Marek~\cite{heule2017solving}
call them the ``best solution in most cases''
for performing large combinatorial searches.
Recently SAT solvers have been spectacularly applied to a number of
long-standing mathematical problems including
the Erd\H os discrepancy conjecture (open for 80 years)~\cite{konev2015computer},
the Boolean Pythagorean triples conjecture (open for 30 years)~\cite{heule2017solving},
and the determination of the fifth Schur number (open for 100 years)~\cite{heule2018schur}.
We briefly outline how SAT solvers were successful on these problems
in Section~\ref{sec:sat}.

Despite these successes, SAT solvers are known to not perform well
for all kinds of combinatorial searches such as those
that require advanced mathematics.  For example,
Arunachalam and Kotsireas~\cite{arunachalam2016hard}
have shown that searching for mathematical
objects defined by autocorrelation
relationships are hard for current SAT solvers.
Similarly, Van Gelder and Spence~\cite{van2010zero} have shown
that proving the nonexistence of certain combinatorial designs
(even some that have intuitively very easy nonexistence proofs)
produce small but very difficult instances for SAT solvers.

\paragraph{Symbolic computation}
Symbolic computation or computer algebra is the branch of computer science concerned
with manipulating algebraic expressions and other mathematical objects.
It has been studied for over sixty years and its successes has lead
to the development of computer algebra systems (CASs) that can
now automatically solve many theoretical and practical mathematical
problems of interest.
For example, a modern computer algebra system
has functionality for things such as
Gr\"obner bases, cylindrical algebraic decomposition,
lattice basis reduction, linear system solving,
arbitrary and high precision arithmetic,
interval arithmetic, linear and nonlinear optimization,
Fourier transforms, Diophantine solving, 
computing automorphism groups, graph algorithms like
determining if a graph has a Hamiltonian cycle, and many other basic
operations like computing the derivative of a function.

Computer algebra is widely used in engineering and science.  For example,
the 1999 Nobel prize in physics was awarded to Gerardus 't Hooft
and Martinus J.~G.~Veltman for using computer algebra to place
particle physics on ``a firmer mathematical foundation''.
Computer algebra has also been used to resolve a number of
long-standing mathematical conjectures.  Three well-known
examples of this are the alternating sign matrix conjecture
(open for 15 years)~\cite{zeilberger1996proof},
the Mertens conjecture (open for 100 years)~\cite{odlyzko1985disproof},
and the Kepler conjecture (open for nearly 400 years)~\cite{lagarias2011kepler}.
We briefly discuss how computer algebra
was used to solve them in Section~\ref{sec:cas}.

Despite these successes, computer algebra systems are not optimized
for all types of problems.  In particular, they are typically
not optimized to perform the kind of general-purpose search with
learning that SAT solvers excel at.  In other words, problems
that require searching through a large combinatorial space
will probably not be solved most effectively
by a computer algebra system.

\paragraph{The best of both worlds}
In this paper we overview the new ``SAT+CAS''
paradigm that harnesses the search power of SAT solvers and the
mathematical abilities of CASs.
This approach provides the best aspects of both the SAT and
CAS approaches while minimizing the weaknesses of each respective
tool.  For example, one of the primary drawbacks of SAT solvers
is that they lack mathematical expressiveness---many mathematical concepts
are difficult or even impossible
to efficiently encode in Boolean logic.
On the other hand, a huge variety of mathematical concepts
can easily be expressed in a CAS.  Thus, the SAT+CAS paradigm
combines the search power of a SAT solver with the expressive
power of a CAS.

\begin{figure}
\begin{center}
\begin{tikzpicture}[align=center,node distance=2.5em]
\node(input){\clap{SAT encoding that Williamson}\\{matrices of order $n$ exist}};
\node[below=of input,text width=5.5em,minimum height=3em,rectangle,draw](sat){Split into subproblems};
\node[node distance=8em,right=of sat,text width=5.5em,minimum height=3em,rectangle,draw](cas){CAS};
\node[below=of sat,text width=5.5em,minimum height=3em,rectangle,draw](sat2){SAT solver};
\node[node distance=8em,right=of sat2,text width=5.5em,minimum height=3em,rectangle,draw](cas2){CAS};
\node[below=of sat2,text width=10em](output){{Williamson matrices}\\{or counterexample}};
\draw[->](input)--(sat);
\draw[->](sat)--(sat2);
\draw[->](sat2)--(output);
\draw[->,transform canvas={yshift=0.6em}](sat)--node[above,text width=6em]{\footnotesize\clap{SAT instances}}(cas);
\draw[<-,transform canvas={yshift=-0.6em}](sat)--node[below,text width=6em]{\footnotesize\clap{inequivalent instances}}(cas);
\draw[->,transform canvas={yshift=0.6em}](sat2)--node[above,text width=6em]{\footnotesize\clap{partial satisfying}\\assignment\\}(cas2);
\draw[<-,transform canvas={yshift=-0.6em}](sat2)--node[below,text width=6em]{\footnotesize\clap{conflict clause}}(cas2);
\end{tikzpicture}
\end{center}
\caption{A diagram outlining how the SAT+CAS paradigm
is applied to the Williamson conjecture.}\label{fig:sat+cas}
\end{figure}
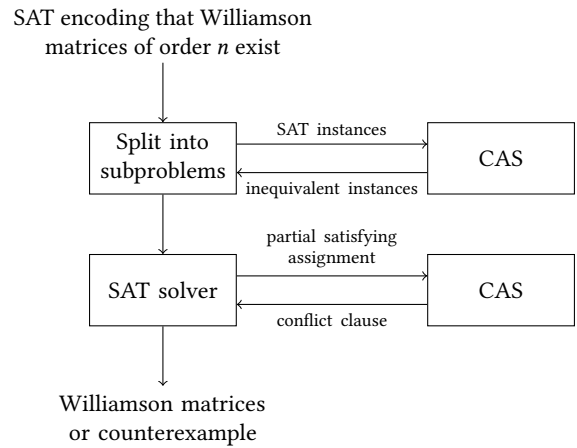

Recently the SAT+CAS paradigm has been used to make progress
on a number of conjectures from combinatorics, graph theory,
and number theory.  In particular, it has
verified a conjecture of Craigen, Holzmann, and Kharaghani,
found three new counterexamples to the good matrix conjecture,
verified the smallest counterexample of the Williamson conjecture,
and is responsible for the current best known results
in the even Williamson, Ruskey--Savage, Norine,
and best matrix conjectures.
We give an overview of these conjectures and how our SAT+CAS
system MathCheck (available at \href{https://uwaterloo.ca/mathcheck}{\nolinkurl{uwaterloo.ca/mathcheck}})
was used to produce these results in Section~\ref{sec:sat+cas}.
A high-level diagram of how MathCheck combines SAT solvers with CASs is shown in Figure~\ref{fig:sat+cas}.
We also briefly discuss how Heule, Kauers, and Seidl have recently used the
SAT+CAS paradigm to find numerous new ways of multiplying $3\times3$
matrices~\cite{heule2019new}.
Finally, we summarize
the kinds of problems for which individually the SAT and CAS paradigms
are insufficient but for which
the SAT+CAS paradigm has been successful in Section~\ref{sec:conclusion}.

\section{Prior Work}\label{sec:prevwork}

In this section we overview the fields of satisfiability checking, symbolic computation,
and the kinds of conjectures resolved using the tools of these fields.
As we will see, these fields have been applied to resolve an impressive variety of conjectures.
Satisfiability checking is particularly good at solving conjectures that can be expressed
only using simple constraints but require an enormous search, while symbolic computation
is particularly good at solving conjectures that require a lot of
complicated mathematical calculations but not a lot of search.

\subsection{SAT solving}\label{sec:sat}

The techniques developed by the field of satisfiability checking
has recently allowed SAT solvers to resolve mathematical conjectures
requiring enormous searches.  In this section we discuss
three of these conjectures.

\paragraph{Erd\H os discrepancy conjecture}
In the 1930s, the prolific mathematician Paul Erd\H os
conjectured that for any infinite $\{\pm1\}$-sequence $X=(x_1,x_2,\dotsc)$
the quantity $D_X(n,k)\coloneqq\abs[\big]{\sum_{i=1}^{n}x_{ki}}$
can be made arbitrarily large by choosing appropriate~$n$ and~$k$.
In 2010, the Polymath project studied the conjecture
and discovered many sequences~$X$ of length 1124 with $D_X(n,k)$ at most~$2$
for all choices of~$n$ and~$k$ for which this quantity is defined.
The sequences were found using a custom computer program and despite
expending a lot of computing effort no longer sequences with this property
were found.  Fields medalist Timothy Gowers would later say
``That was enough to convince me that 1124 was the correct bound
[for the length of sequences~$X$ with $D_X(n,k)$ at most~$2$].''

In 2014, Konev and Lisitsa~\cite{konev2015computer} showed that 1124 was not the
correct bound by using a SAT solver to find a sequence of
length 1160 with $D_X(n,k)$ at most~$2$ for all~$n$ and~$k$.  Furthermore,
they showed that such a sequence of length 1161 could not exist,
thereby resolving the smallest open case of the
Erd\H os discrepancy conjecture.  The full conjecture was resolved
the next year by Terence Tao~\cite{tao2016erdos}, building on results of the Polymath project.

\paragraph{Boolean Pythagorean triples conjecture}
In the 1980s, mathematician Ronald Graham offered a \$100 prize for an answer
to the Boolean Pythagorean triples problem:
Is it possible to split the natural numbers $\{1,2,\dotsc\}$ into two parts
so that all triples $(a,b,c)$ with $a^2+b^2=c^2$ are separated?
In 2008, Cooper and Poirel~\cite{cooper2008pythagorean} found a partition of the natural numbers up to 1344
into two parts with no Pythagorean triple in the same part---this required a custom
computer program and hundreds of hours of computing time.  

In 2016, Heule, Kullmann, and Marek~\cite{heule2017solving} used a SAT solver to find a partition of
the natural numbers up to 7824 into two parts that separated all Pythagorean triples.
Furthermore, they showed that it was not possible to improve this bound---%
there is no 2-partition of the natural numbers up to 7825 that separates
all Pythagorean triples.  The proof found by the SAT solver
was over 200 terabytes and was verified in about 4 CPU years.
Ronald Graham accepted this as a resolution of the Boolean
Pythagorean triples conjecture and awarded his \$100 prize.

\paragraph{Schur number five}
In the 1910s, Issai Schur~\cite{schur1917kongruenz} proved that
for any $k\geq1$ there
exists a largest set $\{1,\dotsc,m\}$ that
can be partitioned into~$k$ parts
such that all triples $(a,b,c)$ with $a+b=c$ are separated.
The value of~$m$ in the above is known as the \emph{Schur number} $S(k)$.
It is possible to check that $S(1)=1$, $S(2)=4$, $S(3)=13$ by hand,
and Baumert and Golomb~\cite{golomb1965backtrack} showed that $S(4)=44$ by a computer search in 1965.
Furthermore, Exoo~\cite{exoo1994lower} showed that $S(5)\geq160$ in 1994
using a combinatorial optimization algorithm.

In 2017, Heule~\cite{heule2018schur} used a SAT solver to show that any partition of
$\{1,\dotsc,161\}$ into 5 parts will not separate all triples
$(a,b,c)$ with $a+b=c$ and therefore showed that $S(5)=160$.
The proof produced by the SAT solver was two petabytes in size
and was verified by a formally-verified proof checker using
about 36 CPU years.

\subsection{Computer algebra}\label{sec:cas}

The techniques developed in the field of computer algebra have been applied
to a huge number of engineering, scientific, and mathematical problems.
In this section we discuss three conjectures where techniques from computer algebra
were essential in the resolution of the conjecture.

\paragraph{Mertens conjecture}
In 1885, Thomas Stieltjes conjectured
(and later independently by F. Mertens) what is
now known as the Mertens conjecture.
The Mertens function is defined by $M(x)\coloneqq\sum_{n\leq x}\mu(n)$
where $\mu(n)\coloneqq(-1)^k$ if the prime factorization of~$n$
consists of~$k$ distinct prime factors
and $\mu(n)\coloneqq0$ if a prime factor appears more than once in the
prime factorization of~$n$.
The Mertens conjecture is that $\abs{M(x)}<\sqrt{x}$ for all~$x>1$.
In the 1970s, the Mertens conjecture was shown to hold for all $x\leq7.8\cdot10^9$.

In 1985, Odlyzko and te Riele~\cite{odlyzko1985disproof} showed that the Mertens conjecture was false.
Their method used lattice basis reduction and arbitrary-precision arithmetic
from the Brent MP package.
The smallest counterexample is still unknown but it is known to be larger than $10^{14}$ and
smaller than $\exp(1.59\cdot10^{40})$.

\paragraph{Alternating sign matrix conjecture}
In the 1980s, Mills, Robbins, and Rumsey~\cite{mills1983alternating} studied alternating sign matrices---%
square $\{0,\pm1\}$-matrices whose rows and columns sum to $1$
and whose nonzero entries alternate sign in each row and column.
They noticed that the number of alternating sign matrices of order~$n\leq10$
was $\prod_{k=0}^n(3k+1)!/(n+k)!$ and conjectured that this
relationship held for all~$n$.

The conjecture was proven by Doron Zeilberger~\cite{zeilberger1996proof}
in the 1990s, crucially relying on the combinatorial functions of the computer
algebra system Maple.  In fact, a Maple package was distributed with
the paper that empirically (and in some cases rigorously) verified
every nontrivial fact in the paper.

\paragraph{Kepler conjecture}
In 1661, the astronomer and mathematician Johannes Kepler conjectured
that the most efficient way of packing spheres in three dimensions is
to stack them in a pyramid shape.  It was still unsolved in 1900
and David Hilbert included it in his famous list of unsolved problems.

In 1998, the mathematician Thomas Hales and his student
Samuel Ferguson~\cite{lagarias2011kepler} proved the Kepler conjecture
using a variety of tools such as global optimization, linear programming,
and interval arithmetic.  Many of the computations
in the proof were performed using Mathematica's arbitrary-precision
arithmetic and double-checked using Maple.  Because of the complexity
of the calculations a team of at least thirteen referees could not be certain
of the proof's correctness after four years.  This lead Hales to
start a project to complete a formal verification of the proof; it
completed in 2014 after a decade of work~\cite{hales2017formal}.

\section{SAT+CAS Paradigm}\label{sec:sat+cas}

As we saw in Section~\ref{sec:prevwork}, the satisfiability checking
and symbolic computation approaches have been applied to resolve a variety of mathematical
conjectures---but each approach has its own advantages and disadvantages.
On the one hand, satisfiability checking is good at solving problems with
enormous search spaces and simple constraints.
On the other hand, symbolic computation is good at solving problems with
sophisticated mathematical calculations.

When a search space becomes too large the overhead associated with
a computer algebra system becomes more pronounced, necessitating the usage
of a more efficient solver.
Currently, SAT solvers are probably the best tools currently available for general purpose
search; they are very difficult to beat because of the decades of engineering
effort that has been aimed at making them efficient.

Given this, Zulkoski, Ganesh, and Czarnecki in 2015 proposed~\cite{zulkoski2015mathcheck} (and independently
by \'Abrah\'am~\cite{abraham2015building}) the \emph{SAT+CAS paradigm} of combining SAT solvers and
CASs to solve conjectures that require both
efficient search and advanced mathematics.  In this section we overview
and explain the major successes of the SAT+CAS paradigm over the last four years.

\subsection{Williamson conjecture}\label{sec:Williamson}

In 1944, the mathematician J. Williamson studied the
Hadamard conjecture from combinatorial design theory.  This conjecture says that
square $\{\pm1\}$-matrices with
with pairwise orthogonal rows exist in all orders $4n$.  He defined a new
class of matrices now known as Williamson matrices that he used
to construct Hadamard matrices of order~$4n$ for certain small values
of~$n$.  Symmetric $\{\pm1\}$-matrices $A$, $B$, $C$, $D$
form a \emph{set of Williamson matrices} (each individual matrix itself
being \emph{Williamson}) if they
are circulant (each row is a cyclic shift of the previous row)
and if $A^2+B^2+C^2+D^2$ is the scalar matrix $4nI$.
It was once considered likely that Williamson matrices exist for all~$n$
and therefore Williamson matrices could provide a route to proving the
Hadamard conjecture~\cite{golomb1963search}.  The conjecture that
Williamson matrices exist in all orders~$n$ has since become known
as the Williamson conjecture.

The hopes that Williamson matrices exist in all orders
were dashed in 1993, when D. \v Z. \DJ okovi\'c~\cite{dokovic1993williamson}
showed that Williamson matrices of order~35 do not exist
by an exhaustive computer search.  \DJ okovi\'c noted that this was
the smallest \emph{odd} counterexample of the Williamson conjecture
but did not specify if it was truly the smallest counterexample.
In 2006, Kotsireas and Koukouvinos~\cite{kotsireas2006constructions}
found no counterexamples in the even orders $n\leq 22$
using the CodeGeneration package
of the computer algebra system Maple.  In 2016,
using an off-the-shelf SAT solver, Bright et~al.~\cite{bright2016mathcheck} found
no counterexamples in the even orders $n\leq30$.
Despite these successes, both the SAT-only and CAS-only
approaches failed to find the smallest counterexample of the Williamson conjecture.

Not only did the SAT+CAS approach successfully find the smallest counterexample, it blew 
the other approaches out of the water by exhaustively solving all even orders up to
seventy~\cite{bright2018sat,bright2019applying}.
The search space up to order $70$ is an astronomical \emph{twenty-five orders of magnitude}
larger than the search space up to order $30$ because
the search space for Williamson matrices grows exponentially in~$n$.
Williamson matrices were found to exist in all even orders $n\leq70$, leading
to the \emph{even Williamson} conjecture that Williamson matrices exist
in all even orders.

The SAT+CAS approach is able to search such large spaces
by exploiting mathematical properties of Williamson matrices that dramatically shrink the search space.
In particular, the most important known filtering property is the \emph{power spectral density (PSD) criterion} that
says that if $A$ is a Williamson matrix of order $n$
with first row $[a_0,\dotsc,a_{n-1}]$ then
\[ \PSD_A(k) \coloneqq \abs[\Big]{\sum_{j=0}^{n-1}a_j e^{2\pi ijk/n}}^2 \leq 4n \]
for all integers~$k$.
This is an extremely strong filtering condition;
a random circulant and symmetric $\{\pm1\}$-matrix $A$ will almost certainty
fail it.  Thus, a solver that is able to effectively exploit the PSD criterion will easily
outperform a solver that does not know about this property.
However, to effectively use it we need
\begin{enumerate}
\item an efficient method of computing the PSD values; and
\item an efficient method of searching while avoiding matrices that fail the filtering criteria.
\end{enumerate}
The fundamental reason for the success of the SAT+CAS paradigm
in regard to the Williamson and even Williamson conjectures is that
CASs excel at~(1) and SAT solvers excel at~(2).

The manner in which the SAT and CAS are combined is demonstrated in Figure~\ref{fig:sat+cas}.
As the SAT solver completes its search it sends to a CAS the matrices $A$, $B$, $C$,
$D$ from partial solutions of the SAT instance.  The CAS then ensures that the matrices
pass the PSD criterion.  If a matrix fails the PSD criterion then a \emph{conflict clause}
is generated encoding that fact.  The SAT solver adds the conflict clause into its
learned clause database, thereby blocking the matrix from being considered in the future.

The search was also parallelized by splitting the search space into many independent subspaces.  Each subspace
had a separate SAT instance generated for it and the SAT instances were solved in parallel.  The CAS was also
useful in the splitting phase by removing instances that were found to be equivalent to other instances
under the known equivalence operations of Williamson matrices.

In the end, our SAT+CAS system MathCheck found over 100,000 new sets of Williamson matrices among 
all even orders $n\leq70$, a new set of Williamson matrices in the odd order~$63$,
and verified that $n=35$ is the smallest counterexample of the Williamson conjecture.

\subsection{Good and best matrix conjectures}

Many variants of Williamson matrices exist; two variants are known as \emph{good matrices}
(introduced by J. Seberry Wallis~\cite{wallis1970combinatorial})
and \emph{best matrices} (introduced by Georgiou, Koukouvinos, and Seberry~\cite{georgiou2001circulant}).
There are several slightly different definitions for such matrices,
but for our purposes we define them
to be circulant matrices $A$, $B$, $C$, $D\in\{\pm1\}^{n\times n}$ that satisfy
$AA^T+BB^T+CC^T+DD^T=4nI$ where $A$ is skew ($A+A^T=2I$) and $D$ is symmetric ($D=D^T$).
Additionally, $B$ and $C$ are skew (for best matrices) or symmetric (for good matrices).

It is known that if good matrices exist of order~$n$ exist then $n$ must be of the form $2r+1$ (i.e., odd)
and if best matrices of order~$n$ exist then $n$ must be of the form $r^2+r+1$.  The good and best matrix
conjectures state that good and best matrices exist in \emph{all} orders of these forms.
In 2002, the good matrix conjecture was shown to hold for all $n\leq39$~\cite{georgiou2002good}
and in 2001 the best matrix conjecture was shown to hold for all $n\leq31$~\cite{georgiou2001circulant}.
In 2018, the best matrix conjecture was shown to hold for all $n\leq43$ and the counterexamples
$n=41$, $47$, and $49$ were found to the good matrix conjecture~\cite{djokovic2018goethals}.

MathCheck has also been applied to the good and best matrix conjectures~\cite{bright2019good,bright2019best}
using a similar method as described in Section~\ref{sec:Williamson} with some encoding adjustments
that are specific to good or best matrices.
For example, if $[d_0,\dotsc,d_{n-1}]$ is the first row of a symmetric best matrix
then it is known that $d_{n/3}=d_0$ when $n$ is a multiple of~$3$.
MathCheck found two new sets of good matrices (for $n=27$ and~$57$) and three new counterexamples
of the good matrix conjecture ($n=51$, $63$, and~$69$).  MathCheck also found
three new sets of best matrices in order $57$ and showed that the best matrix conjecture holds
for all $n\leq57$ (the best currently known result).

\subsection{Craigen--Holzmann--Kharaghani conjecture}

In 2002, Craigen, Holzmann, and Kharaghani~\cite{craigen2002complex} studied
\emph{complex Golay pairs} which are polynomials $f$, $g$ with $\{\pm1,\pm i\}$
coefficients such that $\abs{f(z)}^2+\abs{g(z)}^2$ is constant on the unit circle.
This implies that~$f$ and~$g$ have
the same number of terms and this quantity is known as the
\emph{length} of the polynomial.
Craigen, Holzmann, and Kharaghani performed an exhaustive search for all
complex Golay pairs up to length~$19$ and a partial search up to length~$23$.
They found no complex Golay pairs of length~$23$ and conjectured that they
did not exist.  An exhaustive search was performed by F. Fiedler in
2013~\cite{fiedler2013small} that did not find any complex Golay pairs
of length~$23$, though no implementation was provided 
making it difficult to verify his search.

MathCheck can be used to independently verify the results of Fiedler's
searches~\cite{bright2018enumeration,bright2018complex}.
The first step is to find all single polynomials $f$ that could
appear as a member of a complex Golay pair.  A number of known properties
of complex Golay pairs are used to cut down the search space,
the most important one being that $\abs{f(z)}^2\leq 2n$ where
$n$ is the length of $f$ and $z$ is on the unit circle.

Given a potential $f$ we solve the nonlinear optimization problem
of maximizing $\abs{f(z)}^2$ subject to $\abs{z}=1$ (see
Maple's command \textsc{NLPSolve}) and discard the $f$ whose
maximum is greater than $2n$.  Secondly, we use the known fact that
if $(f,g)$ is a complex Golay pair then
$N_g(s)=-N_f(s)$ for $s=1$, $\dotsc$, $n-1$ where
$N_g$ is the nonperiodic autocorrelation function of $g$.

Once $f$ is known and enough of $g$ is known so that
$N_g(s)\neq-N_f(s)$ can be determined
then a conflict clause is learned blocking the partial
solution from ever being tried again.  This filtering theorem
is very powerful because it often works when only a few coefficients
of $g$ are known.  For example, the SAT solver is able to learn
to never assign both the first and last entries of $g$ to be $1$
at the same time.

\subsection{Ruskey--Savage conjecture}

In 1993, Ruskey and Savage~\cite{ruskey1993hamilton}
asked if every matching (a set of edges without common vertices)
of the hypercube graph with $2^n$ vertices can be extended into
a Hamiltonian cycle of the graph.
In 2007, Fink~\cite{fink2007perfect} noted that this property holds
in the hypercube graphs for $n=2$, $3$, and~$4$ and he proved a weaker form of the
conjecture that he attributes to Kreweras~\cite{kreweras1996matchings}.

In 2015, MathCheck was used to show for the first time that the
Ruskey--Savage conjecture held for the hypercube graph with
$2^5=32$ vertices~\cite{zulkoski2015mathcheck}.  This was accomplished
by using a SAT solver to exhaustively enumerate the matchings of the
hypercube graph and then verifying with a CAS that each matching
extends to a Hamiltonian cycle.  Certain kinds of matchings could
be ignored; for example, the SAT solver
only enumerates maximal matchings (those which cannot be increased
in size while remaining matchings) because if a maximal matching
extends to a Hamiltonian cycle then so do all subsets
of the matching.

Once the CAS verifies that a given matching extends to a Hamiltonian
cycle, a conflict clause is learned that blocks that Hamiltonian cycle (and all subsets
of it) from being considered in the search again.  Furthermore,
it is also effective to have the CAS apply automorphisms of
the hypercube graph to the Hamiltonian cycle it finds
to generate additional Hamiltonian cycles to be blocked~\cite{zulkoski2017combining}.

\subsection{Norine conjecture}

Consider a 2-colouring of the edges of a hypercube graph such
that edges directly opposite each other have opposite colours.
Serguei Norine conjectured that in such a colouring
it is always possible to find two directly opposite vertices
that are joined by a path of edges of a single colour~\cite{norine2008edge}.
In 2013, Feder and Subi reported that the conjecture had been verified for hypercube
graphs with $n=2$, $3$, $4$, and $5$, and proved the conjecture
for a special class of edge colourings~\cite{feder2013hypercube}.

In 2015, MathCheck was used to show for the first time that the Norine conjecture
held for the hypercube graph with $2^6=64$ vertices~\cite{zulkoski2015mathcheck}.
This was accomplished by using a SAT solver to exhaustively enumerate the edge
colourings for which the conjecture was not already known to hold.

Once an edge colouring was found by the SAT solver it was passed to a CAS to
verify that the colouring contains at least two directly opposite vertices
that are connected by a path of a single colour.  If such vertices do
not exist then this colouring forms a counterexample to the conjecture;
otherwise, a conflict clause is generated that blocks this colouring
from appearing in the search again.
In fact, any colouring that includes the monochromatic path that was found
by the CAS can be blocked, since all such colourings cannot be counterexamples
to the Norine conjecture.
Similar to in our work on the Ruskey--Savage conjecture,
it is also effective to have the CAS apply automorphisms of
the hypercube graph to the path that it finds
to generate additional colourings to be blocked~\cite{zulkoski2017combining}.

\subsection{3 by 3 matrix multiplication}

The classical way of multiplying two $2\times2$ matrices uses eight
scalar multiplications; in 1969, Strassen discovered a way to do it
using just seven scalar multiplications~\cite{strassen1969gaussian}.
Two years later, Winograd showed that
it is not possible to do it with six multiplications~\cite{winograd1971multiplication}
and de Groot~\cite{de1978varieties} showed there is essentially one
optimal algorithm.

The optimal algorithm for multiplying $3\times3$ matrices is still unknown
and the best known algorithm uses 23 multiplications~\cite{laderman1976noncommutative}.
Previously, four inequivalent algorithms were known with this complexity.
Recently, Heule, Kauers, and Seidl~\cite{heule2019new} found
over 13,000 additional inequivalent algorithms that use 23 multiplications.
This was achieved using the SAT+CAS paradigm in a multistage process.

In the first stage, they reduce the problem of finding a matrix multiplication
algorithm using 23 scalar multiplications to solving $3^6=729$ cubic equations
in $23\cdot 3^3=621$ variables.
A SAT instance is generated from these equations by reducing them modulo~$2$.
A solution of the SAT instance
then provides a way to multiply $3\times3$ matrices over the
finite field $F_2=\{0,1\}$.

By using various simplifications they found over 270,000 solutions of the
SAT instance.  They then used the computer algebra system Mathematica to determine
that over 13,000 of those solutions are inequivalent.  Finally, they use a
Gr\"obner basis calculation in the computer algebra system Singular to
lift the solutions found for the field $F_2$ to an arbitrary ring.
They report that a small number of solutions over $F_2$ cannot be lifted
in such a way but in most cases each solution provides a new $3\times3$
matrix multiplication algorithm that works in any ring.
None of the algorithms they found could be simplified to use only $22$
multiplications making it tempting to conjecture that such an algorithm
does not exist.

\section{Conclusion}\label{sec:conclusion}

In this article we have surveyed the SAT+CAS paradigm of combining
SAT solvers and computer algebra systems aimed at resolving
mathematical conjectures. It is illuminating to contrast the kind of problems that have been solved
by the SAT and CAS paradigms individually, with those that have been
solved by the combined SAT+CAS paradigm.

We discussed three long-standing mathematical problems in Section~\ref{sec:sat}
for which SAT solvers have been used.
For each problem, attempts to use custom-purpose search code or optimization methods
ultimately proved to not be as successful as using a SAT solver. This is due to the many efficient search heuristics that have been incorporated in modern solvers, as well as the years of refinements that have gone into these solvers. These heuristics have broad applicability for problems from diverse domains.

Additionally, we saw three long-standing conjectures in Section~\ref{sec:cas}
that CAS methods were used to resolve.  In each case, very efficient mathematical
calculations were necessary but efficient search routines were not
the bottleneck in the solutions.  These conjectures would not be a
good fit for SAT solvers because these problems do not admit
natural encodings into Boolean logic.

Note that the eight conjectures from Section~\ref{sec:sat+cas} would be
difficult to resolve using \emph{either} SAT solvers or CASs alone.
In each case, the problems have both a significant search component
(an exponentially growing search space) and a significant mathematical
component (e.g., requiring knowledge of the power spectral density of a circulant
matrix or the automorphism group of a graph).  As we've seen, the SAT+CAS paradigm
is effective at pushing the state-of-the-art in such conjectures.
Simply put, the SAT+CAS paradigm allows the mathematician to solve problems 
that have search spaces too large for CASs and
require mathematical calculations too sophisticated for SAT solvers.

\bibliographystyle{abbrv}
\bibliography{cascon}

\end{document}